\documentclass[twocolumn,showpacs,preprintnumbers,amsmath,amssymb]{revtex4}

\usepackage{graphicx}
\usepackage{dcolumn}
\usepackage{bm}

\usepackage{amsfonts,fancybox,ascmac,amsmath, color}
\usepackage{amsthm}
\usepackage{amssymb}

\theoremstyle{definition}

\newcommand{\mean}[1]{\langle{#1}\rangle}
\newcommand{\bra}[1]{\langle{#1}|}
\newcommand{\ket}[1]{|{#1}\rangle}

\newcommand{\half}{\frac{1}{2}}

\begin{document}


\title{Deterministic generation of Gaussian pure state 
in quasilocal dissipative system}


\author{Yusuke Ikeda}
\author{Naoki Yamamoto}
\affiliation{%
Department of Applied Physics and Physico-Informatics, 
Keio University, Yokohama 223-8522, Japan 
}%




\date{\today}


\begin{abstract}

This paper shows that an arbitrary Gaussian pure state can be 
deterministically generated in a dissipative open system that 
has quasilocal interactions between the subsystems and couples 
to the surrounding environment in a local manner. 
A quasilocal interaction, which means that the interaction 
occurs among only a few subsystems, is a crucial requirement for 
practical engineering of a dissipative system. 
The key idea is that first an auxiliary system having a local 
interaction with the environment is prepared and then  
that auxiliary system is coupled to the underlying target system 
via a set of two-body Hamiltonians in such a way that 
a desired pure state is generated. 
Moreover, we show that even with a simple single-mode auxiliary 
system, the deterministic generation of an arbitrary approximate 
Gaussian cluster state is possible, by devising an appropriate 
switching scheme. 
We discuss in a specific example how much a 
dissipation-induced pure Gaussian state can be perturbed by 
decoherence and parameter error. 

\end{abstract}

\pacs{03.65.Yz, 42.50.Dv}
\maketitle



\section{Introduction}

Preparing a desired pure state is a crucial task in quantum 
information technologies. 
However, in a realistic situation any quantum system unavoidably 
interacts with the surrounding environment and is often described by the 
following Markovian master equation:
\begin{align}
\label{master}
    \frac{d}{dt}\rho(t) =&-i[H, \rho(t)]
\nonumber \\
     +\sum_{i=1}^m &
        \left( L_i\rho(t) L_i^\dagger -\frac{1}{2} L_i^\dagger L_i \rho(t) 
                  -\frac{1}{2}\rho(t) L_i^\dagger L_i \right), 
\end{align}
where $H$ is the system Hamiltonian and $L_i$ is the coupling 
operator representing the interaction between the system and the 
$i$th environment channel. 
Usually the state $\rho(t)$ in Eq.~\eqref{master} dissipatively 
evolves in time towards a mixed state and never recover its purity. 
However, it has been shown in several papers 
\cite{Poyatos1996,Yamamoto2005,Ticozzi2008,Ticozzi2009,
Ticozzi2012,Schirmer2010,Koga} that by engineering suitable 
pairs of the system operators $H$ and $L_i$, the state $\rho(t)$ 
can be uniquely moved to a pure steady state. 
This means that a desired pure state can be deterministically 
generated without specific initialization of the system. 
We actually find some applications of this environment 
engineering approach to entangled state generation 
\cite{Parkins,Kraus,Li2009,Diehl,Mundarain,Wang2010,Rafiee,
Krauter,Muschik} and further some advanced quantum information 
processing such as quantum computation \cite{Verstraete}, 
memories \cite{Pastawski}, and distillation \cite{Vollbrecht}.

\begin{figure}[tb]
 \begin{center}
  \includegraphics[width=8cm]{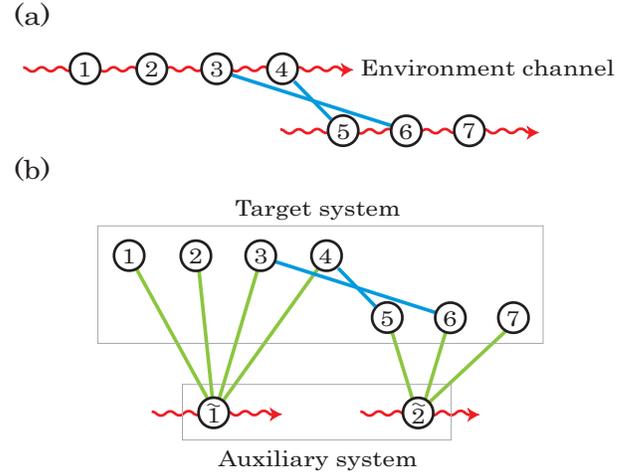}
 \end{center} 
 \caption{
 (a) Example of a quantum oscillator network having a non-quasilocal 
 interaction with the environment. 
 Here $i$ denotes the $i$th oscillator. 
 (b) Extended system that yields the same pure Gaussian 
 steady state as that produced by the system in (a). 
 The auxiliary system couples to the environment locally, 
 whereas it couples to the target system in a quasilocal manner 
 through a certain Hamiltonian. }
 \label{fig:cartoon1}
\end{figure}

Here we mention the quasilocal interaction. 
An operator is called quasilocal if it acts only on a given 
site of the system and its neighborhood; 
particularly in this paper we call the interaction quasilocal 
if it occurs between only two nodes. 
In the case of a quantum oscillator network whose $i$th node 
is a single-mode oscillator of variable $x_i=(q_i, p_i)$, 
an example of a quasilocal coupling is $L=q_1+q_2$. 
However, to generate a desired pure state such as a highly 
entangled state via the above-mentioned dissipation-based method, 
the system is often required to have non-quasilocal (global) 
interaction with the environment. 
Let us consider a seven-node quantum oscillator network depicted in 
Fig.~1(a).
This network has internal couplings among the nodes $x_3$-$x_6$ 
and $x_4$-$x_5$ through a Hamiltonian $H$ in addition 
two global couplings of the nodes $(x_1, x_2, x_3, x_4)$ and 
$(x_5, x_6, x_7)$ to the environments, the operators of which 
are, for instance, $L_1=q_1+q_2+q_3+q_4$ and $L_2=p_5+p_6+p_7$. 
From a practical viewpoint, clearly this kind of global 
interaction is hard to implement. 
Actually, there have been several proposals to engineer a 
dissipative system having a quasilocal interaction with the environment, 
which yet deterministically produces a useful pure state. 
For instance, it was shown in Ref \cite{Kraus} that a qubit two-dimensional  
cluster state can be generated via a quasilocal dissipative 
process and Ref. \cite{Rafiee} demonstrated the case of 
a uniformly distributed entangled state. 
Moreover, for general finite-dimensional systems, Ticozzi and Viola 
provided a condition to determine whether a given pure state can be 
stabilized under fixed locality constraints \cite{Ticozzi2011}. 
However, in the infinite-dimensional case, only a few specific 
results have been reported \cite{Kraus,Koga,Li2009}.

The result obtained in Ref \cite{Koga} is that for any pure Gaussian 
state we can always engineer an open Gaussian system whose unique 
steady state is identical to that pure state. 
However, as mentioned above, this open Gaussian system often has 
to have a global interaction with the environment, such as the system 
shown in Fig.~1(a). 
In this paper, nonetheless, we show that an arbitrary pure Gaussian 
state can be dissipatively generated in a certain extended system 
having only local interaction with the environment. 
The idea is described as follows. 
First, we couple the underlying target system, via a set of two-body 
Hamiltonians, to a certain auxiliary system having local interactions 
with the environment. 
That is, the extended open system composed of the target and the 
auxiliary systems contains only quasilocal interactions 
[see Fig.~1~(b)]. 
Then, based on the result of Ref \cite{Koga}, we will show that a 
Gaussian pure state generated in the original system of interest, 
which can include global interactions with the environment, is also 
generated in the above-mentioned extended system. 
This configuration is actually a generalization of the result 
of Ref \cite{Parkins}, where dissipation-induced generation of a 
two-mode squeezed state in atomic ensembles was demonstrated.

In addition, we present a simple schematic that only utilizes a 
single-mode auxiliary system. 
More specifically, even in the case where the above-mentioned 
result requires us to have a multimode auxiliary system, it 
will be shown that only a single-mode auxiliary system can 
serve to achieve the same goal. 
The key idea is the use of a switching scheme of the auxiliary 
system; 
that is, in the case of the system depicted in Fig.~1, a 
single-mode auxiliary system first couples to the subsystems 
$x_1, \ldots, x_4$ and then switches to couple to the 
subsystems $x_5, x_6, x_7$. 
A particularly important result is that, with this switching 
scheme, an arbitrary approximate Gaussian cluster state 
\cite{Zhang2006,Menicucci2006,Loock,Menicucci2011,Furusawa} 
can be generated dissipatively. 
This is a generalization of the switching scheme proposed by 
Li~\textit{et al.} in Ref \cite{Li2009}, which demonstrated 
dissipative generation of several types of four-mode cluster 
states in an actual physical setup.

Finally, we examine how much a specific dissipation-induced 
pure state is robust against some perturbation. 
The system is a pair of two atomic ensembles that dissipatively 
generates a two-mode squeezed state \cite{Parkins}; here 
damping decoherence and parameter uncertainty are further taken 
into account. 
Then we find the optimal system parameters that maximize the 
entanglement between the atomic ensembles. 
A point worth noting is that these optimal parameters are out of the range 
where the auxiliary cavity mode can be adiabatically eliminated; 
that is, this is an example where the extended system is really 
robust against some perturbation compared to only the target 
system obtained through the adiabatic elimination.

{\it Notation}. 
Let $X=(X_{ij})$ be a matrix whose entry $X_{ij}$ is an operator 
on a Hilbert space or a complex number. 
Then $X^\top=(X_{ji})$ denotes the transpose of $X$. 
Also, $X^\dagger$ means the Hermitian conjugate of $X$; i.e., 
$X^\dagger=(X_{ji}^\dagger)$ for an operator $X_{ij}$ and 
$X^\dagger=(X_{ji}^\ast)$ for a complex number $X_{ij}$. 
We use $\Re(X)$ and $\Im(X)$ to denote the real and imaginary 
parts of $X$. 
For a vector of operators $x$, we define the anti-commutator of 
$x$ by 
\begin{align*}
    \{ x,x^\top \} =xx^\top + (xx^\top )^\top. 
\end{align*}
%


\section{Gaussian dissipative system}

In this section, we describe a general Gaussian dissipative system 
in terms of a quantum stochastic differential equation (QSDE) and 
present the condition for the steady state of this system to be 
pure \cite{Koga}.  The time evolution of a general open quantum system is generated 
by the unitary operator $U(t)$ subjected to the following QSDE 
(in Ito form) \cite{Hudson}:
\begin{align}
\label{QSDE of U}
   dU(t)=\Big\{ \sum_{i=1}^m & 
           \left( L_i dA_i(t)^\dagger-L_i^\dagger dA_i(t) \right)
\nonumber \\
       - &  \left( \sum_{i=1}^m \frac{1}{2}L_i^\dagger L_i+iH \right)dt 
            \Big\} U(t),
\end{align}
with $U(0)=I$ (we have not included the scattering term). 
Here $A_i(t)$ is the annihilation process on the 
$i$th environment vacuum field; 
this satisfies the quantum It\^o rule, e.g., 
$dA_i(t)dA_j(t)^\dagger=\delta_{ij}dt$. 
Let $\rho$ be the initial state of the system and $\ket{0}$ be 
the vacuum state of the whole environment field. 
The master equation \eqref{master} is obtained by 
differentiating $\rho(t)=U(t)(\rho\otimes\ket{0}\bra{0}) U(t)^\dagger$ 
and tracing out the field state.

Next, let $(q_i, p_i)$ be the canonical conjugate pair of the 
$i$th quantum oscillator, which satisfies the canonical 
commutation relation (CCR) $q_ip_j-p_jq_i=i\delta_{ij}$. 
Defining the vector of observables 
$x=[q_1,\dots,q_n,p_1,\dots,p_n]^\top$, we can write the CCR as 
\begin{align*}
   xx^\top -(xx^\top )^\top = i\Sigma_n,~~~
   \Sigma_n
      =\begin{bmatrix} 
           0 & I_n \\
          -I_n & 0 
        \end{bmatrix},
\end{align*}
where $I_n$ denotes the $n \times n$ identity matrix (we often 
drop the subscript $n$). 
A general linear open system can be characterized by the 
following Hamiltonian $H$ and the coupling operators 
$L=[L_1,\ldots,L_m]^\top$: 
\begin{align*}
    H=\frac{1}{2}x^\top Gx, \quad 
    L=Cx,
\end{align*}
where $G=G^\top \in \mathbb{R}^{2n \times 2n}$ and 
$C\in \mathbb{C}^{m\times 2n}$. 
From Eq.~\eqref{QSDE of U}, the QSDE of 
$x(t)=[U(t)^\dagger q_1 U(t), \ldots, U(t)^\dagger q_n U(t), U(t)^\dagger p_1 U(t), \ldots, U(t)^\dagger p_n U(t)]^\top$ 
is given by 
\begin{equation}
\label{linear QSDE}
    dx(t)=Ax(t)dt+BdW(t), 
\end{equation}
where $A=\Sigma_n (G+ \bar{C}^\top \Sigma_m \bar{C}/2)$ and 
$B=\Sigma_n \bar{C}^\top$ with 
$\bar{C}=\sqrt 2[\Re(iC)^\top,\Im(iC)^\top]^\top$. 
Also, we have defined the field operators  
$Q_i=(A_i+A_i^\dagger)/\sqrt{2}$, $P_i=(A_i-A_i^\dagger)/\sqrt{2}i$, 
and $W=[Q_1,\dots,Q_m,P_1,\dots,P_m]^\top$.

A Gaussian system can be fully characterized by only the mean 
vector  $\mean{x}=[\mean{q_1}, \ldots, \mean{q_n}, \mean{p_1}, \ldots, \mean{p_n}]^\top
$ and the 
covariance matrix 
$V=\mean{ \{ \Delta x,\Delta x^\top \} }/2$ 
with $\Delta x=x-\mean{x}$. 
Note that a Gaussian state is pure if and only if 
$1/\sqrt{2^{2n}{\rm det}(V)}=1$. 
The importance of the linear system \eqref{linear QSDE} is in 
the fact that, if the initial state of this system is Gaussian, 
then it preserves the Gaussianity of the state for all time. 
In particular, the covariance matrix $V(t)$ obeys the following 
Lyapunov differential equation: 
\begin{equation}
\label{lyap diff}
    \frac{d}{dt}V(t)=AV(t)+V(t)A^\top +\frac{1}{2}BB^\top.
\end{equation}
Hence the covariance matrix of the steady state is given by 
the solution of the algebraic Lyapunov equation
\begin{align}
\label{lyap}
   AV+V A^\top +\frac{1}{2}BB^\top=0.
\end{align}
This equation has a unique solution if and only if $A$ is a Hurwitz matrix; 
i.e., all eigenvalues of $A$ are in the open left half complex 
plane.

Now we are interested in a linear Gaussian system whose 
steady state is uniquely pure. 
It has been shown in Ref. \cite{Koga} that an arbitrary pure Gaussian 
state can be dissipatively generated, if the system matrices $G$ 
and $C$ can be freely chosen. 
In particular, the following result is useful in this paper:

{\it Theorem 1 \cite{Koga}}.
Suppose that Eq.~\eqref{lyap} has a unique solution $V$. 
Then, this is the covariance matrix of a pure Gaussian state 
if and only if the following matrix equations are satisfied: 
\begin{equation}
\label{Theorem 1}
    \left( V+\frac{i}{2}\Sigma_n \right) C^\top =0,~~~
    \Sigma_n GV+V(\Sigma_n G)^\top =0.
\end{equation}
%


\section{Pure Gaussian state generation via quasilocal 
dissipative environment} 
\label{sec:quasilocal}

To generate a certain desirable pure Gaussian state dissipatively, 
the system is often required to have a global interaction with the 
environment; 
that is, for instance, the coupling operator must be of the 
form $L=q_1+q_2+q_3+\cdots$, which corresponds to $C=(1,1,1,\ldots)$. 
In this section, we show that such a global interaction 
between the system and the environment can always be avoided by 
constructing an auxiliary system that quasilocally couples to 
the target system and locally couples to the environment.


\subsection{The extended dissipative system}
\label{sec:extended}

In addition to the underlying target system consisting of $n$ 
quantum oscillators, we consider an auxiliary system consisting 
of $m$ oscillators. 
Let $x=[q_1,\dots,q_n,p_1,\dots,p_n]^\top$ and 
$\tilde{x}=[\tilde{q}_1,\dots,\tilde{q}_m,\tilde{p}_1,\dots,
\tilde{p}_m]^\top$ be the vectors of canonical conjugate pairs 
of each system. 
We assume that the decoherence of the target system is negligible, 
i.e., the system operators are 
\begin{equation*}
    H=\frac{1}{2}x^\top Gx, \quad 
    L=0,
\end{equation*}
where $G=G^\top \in \mathbb{R}^{2n \times 2n}$. 
The auxiliary system has no self-Hamiltonian and dissipates 
through typical damping channels, hence the system operators 
are given by 
\begin{equation*}
    \tilde{H}=0, \quad 
    \tilde{L}=\sqrt{\kappa}\tilde{a}, 
\end{equation*}
where $\tilde{a}=[\tilde{a}_1,\dots,\tilde{a}_m]^\top$ 
is a vector of annihilation operators, i.e., 
$\tilde{a}_i=(\tilde{q}_i+i\tilde{p}_i)/\sqrt2$. 
The two systems couple via the following 
interaction Hamiltonian:
\begin{align}
\label{Hint1}
     H_{\mathrm{int}}
         =i(\tilde{a}^\dagger Cx-x^\top C^\dagger \tilde{a}),
\end{align}
with $C \in \mathbb{C}^{m \times 2n}$. 
Note that $H_{\mathrm{int}}$ describes a quasilocal interaction 
between the two systems, because it can always be decomposed into 
a sum of two-body Hamiltonians as follows: 
\[
   H_{\mathrm{int}}
    =\sum_{i,j} (x^{(i)\top} K^{(ij)}\tilde{x}^{(j)} 
                  + \tilde{x}^{(i)\top} K^{(ij)\top}x^{(j)}), 
\]
where $x^{(i)}=[q_i, p_i]^\top$ and 
$\tilde{x}^{(i)}=[\tilde{q}_i, \tilde{p}_i]^\top$.  
We will see later that this quasilocal Hamiltonian 
\eqref{Hint1} has the same effect as the  
coupling operator of the system originally of interest, i.e., 
$L=Cx$, when choosing $G$ and $C$ so that they satisfy the condition 
of Theorem~1. 
Defining $\bar{C}=\sqrt 2[\Re(iC)^\top,\Im(iC)^\top]^\top$, we can 
rewrite Eq.~\eqref{Hint1} as 
\begin{equation*}
    H_{\mathrm{int}}
      =\frac{1}{2}( x^\top \bar{C}^\top \tilde{x}
                      +\tilde{x}^\top \bar{C} x ). 
\end{equation*}
Then, the overall system vector 
$\xi(t)=[x(t)^\top, \tilde{x}(t)^\top]^\top$ obeys the following 
linear QSDE: 
\begin{align}
\label{extended system}
    d\xi(t) =A\xi(t) dt + B dW(t),
\end{align}
where 
\begin{equation}
\label{A B matrices}
     A = \begin{bmatrix} 
            \Sigma_n G & \Sigma_n \bar{C}^\top \\ 
            \Sigma_m \bar{C} & -\kappa I_{2m}/2 
         \end{bmatrix}, \quad
     B = -\begin{bmatrix} 
              0 \\ 
              \sqrt\kappa I_{2m} 
          \end{bmatrix}.
\end{equation}
The covariance matrix 
$V(t)= \langle \{ \Delta \xi(t),\Delta \xi(t)^\top  \} \rangle/2$ 
evolves in time through Eq.~\eqref{lyap diff} and its steady 
solution, if it exists, is obtained by solving Eq.~\eqref{lyap}. 
To avoid confusion, we again present the same equation
\begin{align}
\label{fulllyap}
    AV+VA^\top +\frac{1}{2}BB^\top =0, 
\end{align}
where $A$ and $B$ are now given in Eq.~\eqref{A B matrices}.

Recall here that we were originally interested in the system with 
the Hamiltonian $H=x^\top G x/2$ and the coupling operator $L=Cx$, 
which thus obeys the linear dynamics 
\begin{equation}
\label{system of interest}
    dx(t)=A_1x(t)dt+B_1dW(t),
\end{equation}
where $A_1=\Sigma_n(G+ \bar{C}^\top \Sigma_m \bar{C}/2)$ and 
$B_1=\Sigma_n \bar{C}^\top$. 
Note that the dissipation channel $L=Cx$ can be global. 
Its steady covariance matrix in particular is our concern, which 
is subjected to 
\begin{equation}
\label{lyapv1}
     A_1V_1+V_1A_1^\top +\frac{1}{2}B_1B_1^\top =0. 
\end{equation}
The following theorem states that, if the system 
\eqref{system of interest} has a pure steady state, then the 
extended system \eqref{extended system} can dissipatively 
produce the same pure Gaussian state.

{\it Theorem 2}. 
Suppose that Eq.~\eqref{lyapv1} has a unique solution $V_1$ that 
corresponds to a pure Gaussian state. 
Then Eq.~\eqref{fulllyap} has a unique solution 
$V=\mathrm{diag}(V_1,I_{2m}/2)$.

\begin{proof}
First, we prove that Eq.~\eqref{fulllyap} has a unique solution, 
which is equivalent to that $A$ is a Hurwitz matrix. 
For this purpose, let us set $A^\top \eta =\lambda \eta$; 
i.e., $\eta \in \mathbb{C}^{2(n+m)}$ is an eigenvector of $A^\top$ 
and $\lambda \in \mathbb{C}$ is the corresponding eigenvalue. 
Then, multiplying Eq.~\eqref{fulllyap} by $\eta^\dagger$ from the 
left and by $\eta$ from the right, we have 
$\Re(\lambda)=-\eta^\dagger BB^\top\eta/4\eta^\dagger V \eta$. 
This further becomes 
$\Re(\lambda)=-\kappa \eta_2^\dagger \eta_2/4\eta^\dagger V \eta$, 
where we have defined $\eta=[\eta_1^\top,\eta_2^\top]^\top$ with 
$\eta_1\in\mathbb{C}^{2n}$ and $\eta_2\in\mathbb{C}^{2m}$. 
Now let us assume $\eta_2=0$; 
then, from $A^\top\eta=\lambda \eta$ we obtain
\begin{align}
\label{gccond}
    G\Sigma_n^\top \eta_1 =\lambda \eta_1, 
    \quad \bar{C}\Sigma_n^\top \eta_1 =0. 
\end{align}
Moreover, noting the assumption that Eq.~\eqref{lyapv1} has 
a unique solution
\begin{align*}
    V_1=\int_0 ^\infty 
        e^{A_1t} \left( \frac{1}{2}B_1B_1^\top \right) e^{A_1^\top t}dt, 
\end{align*}
we find that $\eta_1$ satisfies 
\begin{align*}
    \eta_1^\dagger V_1 \eta_1 
      = \frac{1}{2} \int_0 ^\infty 
          \|\bar{C}\Sigma_n ^\top e^{A_1^\top t}\eta_1\|^2dt.
\end{align*}
However, this takes zero due to Eq.~\eqref{gccond} and the 
Cayley-Hamilton theorem. 
This conclusion contradicts $V_1>0$, hence we have $\eta_2\neq 0$; 
this further leads to $\Re(\lambda)<0$, implying that $A$ is a Hurwitz matrix.

Next let us prove that $V=\text{diag}(V_1,I_{2m}/2)$ is the solution 
of Eq.~\eqref{fulllyap}. 
When representing $V$ in a block matrix form 
$V=[V_1, V_{12} ; V_{12}^\top, V_2]$, these entries satisfy 
\begin{align}
\label{lyaps}
    & \Sigma_n GV_1+V_1(\Sigma_n G)^\top 
        + \Sigma_n \bar{C}^\top V_{12}^\top 
          + V_{12}(\Sigma_m \bar{C})^\top = 0, 
\nonumber \\
    & V_1(\Sigma_m \bar{C})^\top + \Sigma_n \bar{C}^\top V_2 
        + \Sigma_n GV_{12} - \kappa V_{12}/2 = 0, 
\nonumber \\
    & \Sigma_m \bar{C} V_{12} + V_{12}^\top (\Sigma_m \bar{C})^\top 
        - \kappa V_2 + \kappa I_{2m}/2 = 0. 
\end{align}
Now the assumption is that $V_1$ corresponds to a unique pure 
steady state, hence from Theorem~1 we have 
\begin{align}
\label{cond1}
    V_1(\Sigma_m \bar{C})^\top 
         + \frac{1}{2}\Sigma_n \bar{C}^\top = 0, \quad 
    \Sigma_n GV_1 + V_1(\Sigma_n G)^\top = 0.
\end{align}
Note that the former condition comes from the fact that Theorem~1 
yields $V_1\Re(iC)^\top-\Sigma_n \Im(iC)^\top/2=0$ and 
$V_1\Im(iC)^\top+\Sigma_n\Re(iC)^\top/2$. 
Then, we find that the set of matrices $V_{12}=0$, $V_2=I_{2m}/2$, 
and $V_1$ satisfying Eq.~\eqref{cond1} is the solution to 
Eq.~\eqref{lyaps}.  
Because Eq.~\eqref{lyaps} has a unique solution as shown in the 
former part of this proof, 
$V=\mathrm{diag}(V_1,I_{2m}/2)$ is the unique solution.
\end{proof}

This theorem states that, once we find a suitable pair of matrices 
$G$ and $C$ such that the system \eqref{system of interest} 
dissipatively and uniquely generates a desired pure Gaussian state, 
the same goal can be achieved by alternatively constructing the 
extended system \eqref{extended system}. 
Note again that the system \eqref{system of interest} 
can couple to the environment globally, while the extended 
system \eqref{extended system} locally couples to the 
environment through the operator 
$\tilde{L}_i=\sqrt{\kappa}\tilde{a}_i$ and its internal modes 
quasilocally couple with each other through the Hamiltonian 
$H_{\rm int}$. 
That is, the target system stabilizing a desired pure state 
can be realized as a subsystem of an extended system having local 
coupling to the environment and quasilocal internal couplings 
among the nodes.

{\it Remark 1}.
As an auxiliary system, we usually take a system with very fast 
modes that can be adiabatically eliminated. 
In our case, even when the assumption of Theorem~2 does not hold, 
by taking $\kappa$ sufficiently large, the auxiliary system rapidly 
converges to the vacuum and the mode $\tilde{x}(t)$ can be 
adiabatically eliminated. 
Then, the target system is approximated by the system whose coupling 
operator is $L=2Cx/\sqrt\kappa$ \cite{Nurdin}. 
In this sense, Theorem~2 implies that we can treat the system 
as if the auxiliary modes were heavily damped as long as the pure 
steady state condition is satisfied.


\subsection{Generation of two-mode squeezed state} 
\label{sec:epr}

We consider here a two-mode Gaussian system studied in Ref \cite{Parkins}. 
Although for this system the requirement of quasilocality is 
already satisfied, this example clearly illustrates our idea.

\begin{figure}[tb]
 \begin{center}
  \includegraphics[width=5.5cm]{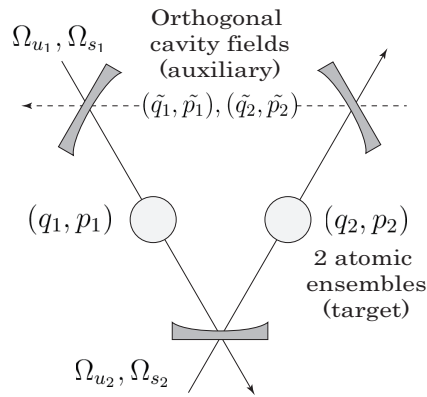}
 \end{center} 
 \caption{
Two atomic ensembles trapped in a two-mode optical cavity. 
Here $\Omega_{u_i}$ and $\Omega_{s_i}$ are the Rabi frequencies 
of the $i$th external laser fields.
 }
 \label{fig:par}
\end{figure}

The physical setup is depicted in Fig.~\ref{fig:par}; 
the target system is two atomic ensembles trapped in a ring-type 
cavity, while the auxiliary system corresponds to this optical 
cavity composed of two propagating modes. 
These two systems interact with each other via external pulse lasers 
with Rabi frequencies $\Omega_{u_i}$ and $\Omega_{s_i}$ $(i=1,2)$. 
Here we assume that the number of atoms in each ensemble is 
sufficiently large; 
then the collective spin component of the atomic ensemble can be 
approximated by an annihilation operator and consequently 
the interaction Hamiltonian is given by 
\begin{align*} 
    H_{\mathrm{int}}=
       &\left \{ \tilde{a}_1^\dagger (\sqrt{N_1}\beta_{u_1}a_1 
           + \sqrt{N_2}\beta_{s_2}a_2^\dagger )+\mathrm{H.c.} \right \} 
\nonumber \\
     + &\left \{ \tilde{a}_2^\dagger (\sqrt{N_1}\beta_{s_1}a_1^\dagger 
           + \sqrt{N_2}\beta_{u_2}a_2 )+\mathrm{H.c.} \right \}, 
\end{align*}
where $a_i$ and $\tilde{a}_i$ are the annihilation operators of 
the $i$th atomic ensemble and the $i$th cavity mode, respectively. 
Here $N_i$ denotes the number of atoms of the $i$th ensemble, and also 
\begin{align*}
   \beta_{u_i}=\frac{\Omega_{u_i}g_{u_i}^\ast }{2\Delta_u}, \quad 
   \beta_{s_i}=\frac{\Omega_{s_i}g_{s_i}^\ast}{2\Delta_s}, \quad i=1,2, 
\end{align*}
where $g_\bullet$ and $\Delta_\bullet$ denote the coupling strength 
and the detuning, respectively. 
If we set the parameters as $N_i=N,~g_{u_i}=g_{s_i}=g~(i=1,2)$, and 
$\Delta_u=\Delta_s=\Delta$, then the interaction Hamiltonian can be 
simply written as
\begin{align*}
   H_{\mathrm{int}}
      =\frac{\sqrt{N}g}{2\Delta} \big[
        &\left \{ \tilde{a}_1^\dagger 
            (\Omega_{u_1}a_1+\Omega_{s_2}a_2^\dagger )
                 +\mathrm{H.c.} \right \} 
\nonumber \\
      + &\left \{ \tilde{a}_2^\dagger 
            (\Omega_{s_1}a_1^\dagger +\Omega_{u_2}a_2 )
               +\mathrm{H.c.} \right \} 
                  \big].
\end{align*}
Further, let us set the Rabi frequencies as 
$\Omega_{u_1}=\Omega_{s_2}=\Omega>0$ and 
$\Omega_{u_2}=\Omega_{s_1}=r\Omega$, 
where $r\in [0,1)$ is a parameter. 
Then, the interaction Hamiltonian is of the form \eqref{Hint1} 
with 
\begin{align}
\label{cepr}
     iC = \frac{\mu}{\sqrt2}
            \begin{bmatrix} 
               1 & r & i & -ir \\ 
               r & 1 & -ir & i 
            \end{bmatrix}, 
\end{align}
where $\mu=\sqrt{N}g\Omega/2\Delta$. 
For $^{87}$Rb atoms, the spontaneous emission of each atom is 
negligible and thus we can set $L=0$, for typical values of the 
parameters, $\mu,\kappa \approx$ 100~kHz and $r=0.8$ \cite{Parkins}. 
Also, the number of atoms is large enough so that the self 
Hamiltonian of the atomic ensembles is assumed to be 0, i.e., $H=0$ 
or equivalently $G=0$.

The coupling operator of the cavity is given by 
$\tilde{L}=\sqrt\kappa \tilde{a}$, where $\kappa$ is the damping 
rate, and we assume that the detuning of the cavity is 0, i.e., 
$\tilde{H}=0$.

Now, for the matrix $C$ given by Eq.~\eqref{cepr} and $G=0$, the 
Lyapunov equation \eqref{lyapv1} has the following unique solution: 
\begin{align*}
    V_1=\frac{1}{2}
          \begin{bmatrix}
               \begin{matrix} 
                  \cosh(2\xi) & -\sinh(2\xi) \\ 
                  -\sinh(2\xi) & \cosh(2\xi) 
               \end{matrix} & 0 \\
               0 & \begin{matrix} 
                      \cosh(2\xi) & \sinh(2\xi) \\ 
                      \sinh(2\xi) & \cosh(2\xi) 
                   \end{matrix}
          \end{bmatrix}, 
\end{align*}
where $\xi=\tanh^{-1}(r)$. 
This is the covariance matrix of a pure two-mode squeezed state. 
Therefore, from Theorem~2, the Lyapunov equation \eqref{fulllyap} 
has a unique solution $V=\mathrm{diag}(V_1,I_4/2)$; 
that is, the pair of atomic ensembles acquires the two-mode 
squeezed state at steady state, while the auxiliary cavity mode 
becomes a trivial coherent state. 
Equivalently, the whole four-mode atom-cavity system generates 
the same atomic steady state as that generated in the two-atomic-
ensemble system with the Hamiltonian $H=0$ and the coupling 
operators 
\begin{align*}
    L_1=\mu (a_1+ra_2^\dagger ), \quad 
    L_2=\mu (a_2+ra_1^\dagger ).
\end{align*}
Another physical realization of this purely dissipative system was 
proposed in Refs. \cite{Krauter,Muschik}.

\section{Dissipation-induced pure cluster state 
with single environment channel} 
\label{sec:switch}

As shown in the preceding section, an arbitrary pure Gaussian state 
can be generated in the target system, by introducing a certain 
$m$-mode auxiliary system that locally couples to $m$ environment 
channels. 
From an engineering viewpoint, it is clearly convenient if we can 
achieve the goal with the auxiliary system having a small number 
of modes $m$. 
In particular, let us consider the simple case $H=0$; 
i.e., the system Hamiltonian $H$ is negligible compared to the 
interaction Hamiltonian, as in the case of Sec.~\ref{sec:epr}. 
In this case, we need the condition $m=n$, which is understood 
from Theorem~1 together with the fact that, for a covariance matrix 
$V$ corresponding to a pure Gaussian state, $V+i\Sigma_n/2$ has 
an $n$-dimensional kernel \cite{Simon}. 
However, this requirement is very demanding especially when 
$n$ is large.

In this section we present a general solution to the above-posed 
problem. 
That is, we prove that even for a general $n$-mode system with 
$H=0$, only a single-mode auxiliary system coupling to a single 
environment channel introduces a dissipative mechanism that drives 
the system state to an arbitrary approximate Gaussian cluster state 
\cite{Zhang2006,Menicucci2006,Loock,Menicucci2011,Furusawa}. 
This result is significant in the sense that a most simple dissipative 
system deterministically generates a most useful quantum state from 
the quantum information viewpoint. 
The key idea is the use of a switching scheme of the interaction 
Hamiltonian between the target and the auxiliary systems, which 
is a generalization of the idea proposed by Li \textit{et al.} 
in Ref. \cite{Li2009}.


\subsection{The switching scheme}
\label{switching scheme}

As introduced above, in this section we consider a most simple 
system; 
that is, the system's Hamiltonian is negligible ($H=0$) 
and the auxiliary system is single-mode ($m=1$), in addition to the 
assumptions $L=\tilde{H}=0$ and $\tilde{L}=\sqrt{\kappa}\tilde{a}$. 
We represent the interaction Hamiltonian \eqref{Hint1} in 
the following form:
\begin{align}
\label{int2}
    H_{\text{int}}
     =\mu\,\sum_{j=1}^n 
         \big\{ \tilde{a}_1^\dagger (\alpha_j a_j +\beta_j a_j^\dagger) 
                 + \text{H.c.} \big\},
\end{align}
where $a_j~(j=1,\dots,n)$ denotes the $j$th annihilation operator 
of the target system. 
$\alpha_j, \beta_j \in \mathbb{C}$ and $\mu \in \mathbb{R}$ are 
parameters.

First let us consider the case $n=1$ and set the parameters as 
$\alpha_1=1$ and $\beta_1=r\in [0,1)$. 
Then, the corresponding $C$ matrix in Eq.~\eqref{Hint1} is given 
by $C=\mu[1+r, i(1-r)]/\sqrt{2}$ and Eq.~\eqref{lyapv1} has the 
following unique solution: 
\begin{align}
\label{one mode squeezed state}
   V_1=\frac{1}{2}
         \begin{bmatrix} 
            e^{-2\xi} & 0 \\ 
            0 & e^{2\xi} 
         \end{bmatrix}, 
\end{align}
where $\xi=\tanh^{-1}(r)$. 
This is the covariance matrix of a pure squeezed state, 
hence from Theorem~2 we find that Eq.~\eqref{fulllyap} has a 
unique solution $V=\mathrm{diag}(V_1,I_2/2)$. 
As a result, when $n=1$, the target system acquires a pure 
squeezed state while the auxiliary state becomes vacuum, 
through the interaction Hamiltonian \eqref{int2}.

Next we consider the case of $n>1$. 
In this case, as mentioned above, Eq.~\eqref{lyapv1} does not 
have a unique solution, hence Theorem~2 cannot be directly applied. 
Now let us consider a unitary matrix $U\in \mathbb{C}^{n\times n}$ 
and the unitary coordinate transformation $a'=Ua$ of the vector of 
system variables $a=[a_1, \dots, a_n]^\top$. 
In terms of the quadratures $q_i'=(a_i'+a_i'\mbox{}^\dagger)/\sqrt{2}$ 
and $p_i'=(a_i'-a_i'\mbox{}^\dagger)/\sqrt{2}i$, we find that $x'=[q'_1, \ldots, q'_n, p'_1, \ldots, p'_n]^\top$ 
is the symplectic and orthogonal transformation of $x$ as follows:
\begin{align}
\label{symplectic}
    x'=Sx, \quad 
    S=\begin{bmatrix} 
        \Re(U) & -\Im(U) \\ 
        \Im(U) & \Re(U) 
      \end{bmatrix}. 
\end{align}
Note that, since $S$ is symplectic, $x'$ satisfies the CCR: 
$x'x'\mbox{}^\top-(x'x'\mbox{}^\top)^\top=iS\Sigma_n S^\top
=i\Sigma_n$.

\begin{figure}[tb]
\begin{center}
\includegraphics[width=8cm]{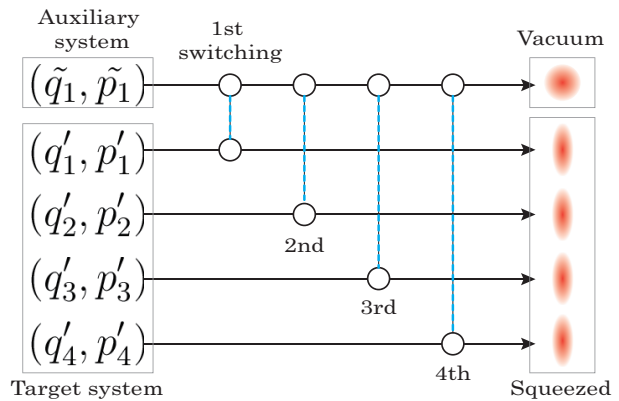}
\end{center} 
\caption{Switching scheme.}
\label{fig:cartoon2}
\end{figure}

We explain here the idea of switching. 
At the $k$th switching stage, the parameters 
$(\alpha_j^{(k)}, \beta_j^{(k)})$ of the interaction Hamiltonian 
\eqref{int2} are chosen as follows:
\begin{equation}
\label{parameter switching law}
     \alpha_j^{(k)} = U_{kj},~~~
     \beta_j^{(k)} = r U^\ast_{kj},~~~
     r\in [0,1), 
\end{equation}
where $U_{kj}$ is the $(k,j)$ element of the unitary matrix 
introduced above. 
With this choice, the interaction Hamiltonian \eqref{int2} is 
written as 
\begin{align}
\label{inti}
    H_{\text{int}}^{(k)}
       =\mu \left \{ \tilde{a}_1^\dagger (a'_k+ra_k'\mbox{}^\dagger ) 
           + \text{H.c.} \right \}.
\end{align}
This is no more than the Hamiltonian discussed in the case $n=1$. 
Hence, in the $k$th switching stage the $k$th mode $(q_k', p_k')$ 
deterministically changes to a pure squeezed state with covariance 
matrix \eqref{one mode squeezed state}. 
Note that during the $k$th stage, the other system variables 
$(q_\ell', p_\ell'),~\ell\neq k$ do not change at all. 
Therefore, applying the interaction Hamiltonian \eqref{inti} 
repeatedly by changing $k=1, 2, \ldots, n$, the corresponding 
system variable $(q_k',p_k')$ gets squeezed in this order, as 
schematically shown in Fig.~\ref{fig:cartoon2}. 
In particular, if the initial state is the ground state 
with respect to $x'$ (and thus $x$ as well), 
all the off-diagonal elements of the covariance matrix remain 0
and then the system's covariance matrix 
$V_1'=\mean{ \{ \Delta x', \Delta x'\mbox{}^\top \} }/2$ becomes 
$V_1'={\rm diag}(e^{-2\xi}I_n, e^{2\xi}I_n)/2$.
As a result, the steady state of the target system of the mode 
$x=S^\top x'$ is a pure Gaussian state with covariance matrix 
\begin{align}
\label{sol1}
    V_1=\frac{1}{2}S^\top 
           \begin{bmatrix} 
              e^{-2\xi}I_n & 0 \\ 
              0 & e^{2\xi}I_n 
           \end{bmatrix}S. 
\end{align}
This is a unitary transformed pure squeezed state; 
we will see later that this state can represent any approximate 
Gaussian cluster state by appropriately choosing the unitary 
matrix $U$ or equivalently the switching parameters 
$(\alpha_j^{(k)}, \beta_j^{(k)})$.

\subsection{Generation of CV cluster state}
\label{cv cluster state}

A Gaussian cluster state is an entangled state of great 
importance, particularly in one-way quantum computation 
\cite{Menicucci2006,Furusawa}. 
Hence, this state should be a target that is dissipatively 
generated with the scheme presented in this paper, particularly 
with the switching scheme. 
Thus we show here how to chose the unitary matrix $U$ so 
that the covariance matrix \eqref{sol1} represents a given 
target Gaussian cluster state.

Here is the definition of a Gaussian cluster state \cite{Menicucci2011}: 
Let $A=A^\top \in \mathbb{R}^{n \times n}$ be the adjacency 
matrix representing the graph structure of a cluster state 
of interest; i.e., the $(i,j)$ element of $A$ represents the 
weight of the coupling between the $i$th and the $j$th nodes 
of the network. 
Then the approximate Gaussian cluster state is defined as a state 
satisfying 
\begin{align}
\label{def of A-graph state}
    \lim_{\alpha \to \infty} \mathrm{cov}(p-Aq)=0,
\end{align}
where $\alpha \in \mathbb{R}$ is a certain parameter contained 
in the state; 
it often corresponds to a squeezing parameter. 
Also we have defined 
$\mathrm{cov}(x)=\langle\{\Delta x,\Delta x^\top \}\rangle/2$. 
It is of course impossible in reality to take the limit 
$\alpha \to \infty$, hence we call the state with finite $\alpha$ 
the approximate Gaussian cluster state.

Now we describe a relation that connects a given adjacency 
matrix $A$ and the unitary matrix $U$ characterizing the 
covariance matrix \eqref{sol1}. 
As will be mentioned later in Remark~2, one such relation 
has already been obtained in Refs. \cite{Furusawa,Loock} particularly 
for the aim of constructing a concrete optical process 
corresponding to $U$, but we here provide an alternative 
method in rather an abstract way.

{\it Proposition 1}.
\label{pro2}
Let $A=A^\top \in \mathbb{R}^{n \times n}$ be the adjacency 
matrix of a given Gaussian cluster state and define $N=-(iI_n+A)$. 
Then the polar decomposition $N=RU$ yields a unitary matrix $U$ 
and a real matrix $R$. 
The Gaussian state with covariance matrix \eqref{sol1} characterized 
by this unitary $U$ then satisfies 
\begin{align} \label{eq prop 1}
     \mathrm{cov}(p-Aq)=\frac{1}{2}(I_n+A^2)e^{-2\xi}, 
\end{align}
hence it is an approximate Gaussian cluster state converging to 
the idel one in the limit of $\xi\rightarrow\infty$.

\begin{proof} 
First, $R$ is real because 
$R=\sqrt{NN^\dagger}=I_n+A^2\in\mathbb{R}^{n\times n}$. 
Next, for the Gaussian state with covariance matrix \eqref{sol1} 
we have 
\begin{align*}
    &\text{cov}(p-Aq)
       =[-A, I_n]\, \half S^T 
            \begin{bmatrix} 
               e^{-2\xi}I_n & 0 \\ 
               0 & e^{2\xi}I_n 
            \end{bmatrix}
          S \begin{bmatrix}
              -A \\ 
              I_n 
            \end{bmatrix} 
\nonumber \\
    & \hspace{1.8cm}
     =\half e^{-2\xi}F_1^\top F_1 
       + \half e^{2\xi}F_2^\top F_2, 
\end{align*}
where $F_1=\Re(U)A+\Im(U)$ and $F_2=\Im(U)A-\Re(U)$. 
Noting that $RU=N$ and $R$ is real, we have $\Re(U)=-R^{-1}A$ and 
$\Im(U)=-R^{-1}$. 
Hence we have $F_1=-R^{-1}(I_n+A^2)$ and $F_2=0$. 
Furthermore, noting that $I_n=UU^\dagger =R^{-1}NN^\dagger R^{-\top}$, 
we have $RR^\top =NN^\dagger =I_n+A^2$, thus 
$F_1^\top F_1=(I_n+A^2)R^{-\top}R^{-1}(I_n+A^2)=I_n+A^2$. 
As a result, we have $\text{cov}(p-Aq)=(I_n+A^2)e^{-2\xi}/2$. 
\end{proof}

The merit of this result is that the unitary matrix $U$ is 
straightforwardly constructed from a given $A$ compared to the 
result of Refs. \cite{Furusawa,Loock}, although in this case $U$ does 
not have a clear correspondence to some optical realizations. 
However, Eq.~\eqref{parameter switching law} clarifies how to 
physically implement the interaction Hamiltonian \eqref{int2}.

{\it Remark 2}.
The approximate Gaussian cluster state with covariance matrix
\eqref{sol1} can be deterministically generated on optical
fields, by the following method \cite{Furusawa,Loock}:
We prepare $n$ independent and identical squeezed light
fields, and mix them via some passive optical devices such
as a beam splitter and a phase shifter in a specific order
determined from the adjacency matrix $A$.
The collection of these transformations is totally represented
by a unitary matrix;
if we denote that unitary matrix as $U^\dagger$, the covariance
matrix of the output fields is identical to Eq.~\eqref{sol1}.
That is, the unitary transform in the switching scheme corresponds
to the scattering process on optical fields, and the
dissipation-induced pure squeezed states correspond to the
initially-prepared optical squeezed states.
Hence, the presented switching scheme can be interpreted as a
dissipative counterpart to the optical scheme proposed in
Refs. \cite{Furusawa,Loock}.
The biggest difference between these two schemes is that in the
dissipative case we consider a state generated in matter whereas
the optical state exists in a flying light field;
the former can be later manipulated or stored, while the latter
is suited for propagating quantum information.
A specific relation between these two regimes was discussed
in Ref. \cite{YamamotoRS2011}.

{\it Remark 3}. 
In Proposition 1 the polar decomposition is used to find the
appropriate unitary matrix $U$, but it is clear from the proof
that only a certain decomposition of the form $N = RU$ with $U$
unitary and $R$ real gives the same relation \eqref{eq prop 1}.
Here we see that, besides the polar decomposition, the Gram
Schmidt procedure also serves as a convenient method to obtain
such a decomposition.
Let $U$ be a unitary matrix whose row vectors are obtained from
the Gram Schmidt procedure of the row vectors of $N$.
Then we immediately have the relation $N = RU$ with a real
lower triangular matrix $R$, which is called the RQ decomposition.
Hence the Gram-Schmidt procedure also yields the unitary matrix
satisfying the condition in Proposition~1.


\subsection{Physical realization}
\label{sec:ring}

\begin{figure}[tb]
\begin{center}
\includegraphics[width=6.0cm]{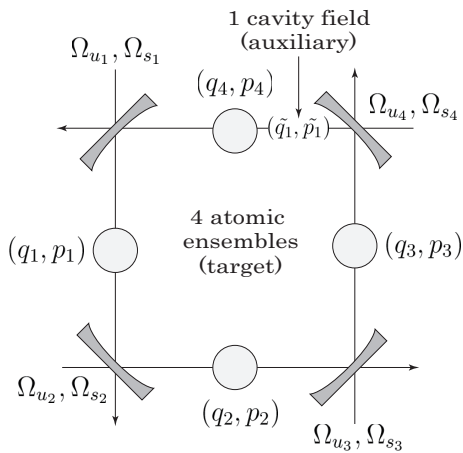}
\end{center} 
\caption{
Atomic ensembles trapped in a single-mode optical cavity. 
Here $\Omega_{u_i}$ and $\Omega_{s_i}$ are the Rabi frequencies 
of the $i$th external laser fields.}
\label{fig:setup2}
\end{figure}

We can implement the presented switching scheme for $n$ atomic 
ensembles (target system) trapped in a single-mode optical 
cavity (auxiliary system) in a similar configuration studied 
in Sec.~\ref{sec:epr}. 
The physical setup in the case $n=4$ is depicted in 
Fig.~\ref{fig:setup2}. 
Let $a_j$ be the annihilation operator approximating the collective 
spin component of the $j$th atomic ensemble and $\tilde{a}_1$  
the annihilation operator of the cavity mode. 
Then the interaction Hamiltonian is given by \cite{Li2009}
\begin{align*} 
    H_{\mathrm{int}}
       =\frac{\sqrt{N}g}{2\Delta}
           \sum_{j=1}^n \Big[ \tilde{a}_1^\dagger 
               (\Omega_{u_j}e^{i\phi_{u_j}}a_j 
      + \Omega_{s_j}e^{i\phi_{s_j}}a_j^\dagger ) + \mathrm{H.c.} 
           \Big], 
\end{align*}
where $\phi_\bullet \in [0,2\pi)$ is the laser phase. 
The switching scheme shown in Sec.~\ref{sec:switch} suggests 
that, at the $k$th switching stage, we choose the parameters as 
\begin{equation}
\label{realization parameter}
    \Omega_{u_j}e^{i\phi_{u_j}}=\Omega \,U_{kj},~~
    \Omega_{s_j}e^{i\phi_{s_j}}=r\Omega \,U^\ast_{kj},~~
    r\in [0,1), 
\end{equation}
where $\Omega>0$ is a parameter. 
The unitary matrix $(U_{kj})$ is determined from the target 
Gaussian cluster state. 
As proven in Sec.~\ref{switching scheme}, the whole state of 
the atomic ensembles deterministically reaches this target, 
if they are all initially set to the ground states.


\subsection{Example}

\begin{figure}[tb]
\begin{center}
\includegraphics[width=2.5cm]{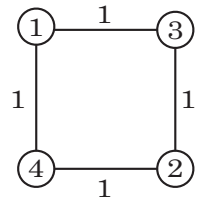}
\end{center} 
\caption{Graphic structure of the four-node square cluster state.}
\label{fig:square}
\end{figure}

Let us consider the problem of dissipatively generating a 
four-node square cluster state, depicted in Fig.~\ref{fig:square}, 
in the atomic ensemble system discussed in the preceding section. 
The connecting edges between the nodes are equally weighted and 
thus the adjacency matrix $A$ of this graph state is given by 
\begin{align*}
     A = \begin{bmatrix} 
            0 & 0 & 1 & 1 \\ 
            0 & 0 & 1 & 1 \\ 
            1 & 1 & 0 & 0 \\ 
            1 & 1 & 0 & 0 
         \end{bmatrix}.
\end{align*}
We follow Proposition~1 to determine a unitary matrix $U$ 
characterizing the switching law \eqref{parameter switching law}. 
In particular, here we utilize the Gram-Schmidt procedure 
(see Remark~3). 
To make the calculation simple, we orthogonalize the following 
matrix: 
\begin{align*}
    N'=
       \begin{bmatrix}
          1 & -1 & 0 & 0 \\
          0 & 0 & 1 & -1 \\
          1 & 0 & 0 & 0 \\
          0 & 0 & 1 & 0 
       \end{bmatrix}N
     =
       \begin{bmatrix}
          i & -i & 0 & 0 \\
          0 & 0 & i & -i \\
          i & 0 & 1 & 1 \\
          1 & 1 & i & 0 
       \end{bmatrix}. 
\end{align*}
Then, $U$ is given by 
\begin{align}
\label{unitary}
    U = \begin{bmatrix}
            -i[1,-1,0,0]/\sqrt{2} \\ 
            -i[0,0,1,-1]/\sqrt{2} \\
            -[i,i,2,2]/\sqrt{10} \\
            \mbox{}[2,2,i,i]/\sqrt{10} 
        \end{bmatrix}.
\end{align}
Actually with this choice we have 
\begin{align}
\label{cov cluster}
      \text{cov}(p-Aq)
         =\frac{1}{2}e^{-2\xi}
             \begin{bmatrix} 
                 3 & 2 & 0 & 0 \\ 
                 2 & 3 & 0 & 0 \\ 
                 0 & 0 & 3 & 2 \\ 
                 0 & 0 & 2 & 3
             \end{bmatrix}, 
\end{align}
thus the state approximates well the target square cluster state 
when large $\xi$ is taken. 

The switching law \eqref{realization parameter} with the 
unitary matrix \eqref{unitary} clarifies how to choose the 
laser parameters as shown below. 
At the first switching stage, they are determined from the 
first row vector of $U$ as 
\begin{align*}
    & \Omega_{u_i}=\frac{\Omega_{s_i}}{r}
                  =\frac{1}{\sqrt{2}}\Omega,~~~
    i=1,2, 
\\ \nonumber
    & \Omega_{u_i}=\Omega_{s_i}=0,~~~
    i=3,4, 
\\ \nonumber
    & \phi_{u_1}=\frac{3}{2}\pi, \quad 
      \phi_{s_1}=\frac{1}{2}\pi, \quad 
    \phi_{u_2}=\frac{1}{2}\pi, \quad \phi_{s_2}=\frac{3}{2}\pi. 
\end{align*}
Through the interaction Hamiltonian with these parameters, 
the CCR pair $(q'_1,p'_1)$ gets squeezed in the long time limit. 
Then we switch the parameters and set: 
\begin{align*}
    & \Omega_{u_i}=\Omega_{s_i}=0, ~~~ i=1,2, 
\\ \nonumber
    & \Omega_{u_i}=\frac{\Omega_{s_i}}{r}
                  =\frac{1}{\sqrt{2}}\Omega, ~~~ i=3,4, 
\\ \nonumber
    & \phi_{u_3}=\frac{3}{2}\pi, 
      \quad \phi_{s_3}=\frac{1}{2}\pi, \quad 
    \phi_{u_4}=\frac{1}{2}\pi, \quad \phi_{s_4}=\frac{3}{2}\pi. 
\end{align*}
Then the second node $(q'_2, p'_2)$ gets squeezed. 
By repeating a similar procedure, each CCR pair $(q'_i,p'_i)$, 
$i=1,\dots,4$ becomes a squeezed state with covariance matrix 
\eqref{one mode squeezed state}. 
As a result, if each atomic ensemble is in the ground state at 
the initial time, the whole state changes to a pure Gaussian state 
with covariance matrix \eqref{sol1}, 
implying that the system reaches the target approximate Gaussian 
cluster state satisfying Eq.~\eqref{cov cluster}.

All of the above results, including the switching law of the 
parameters of the interaction Hamiltonian, have been obtained 
in Ref. \cite{Li2009}; 
here we have shown the same result from a general standpoint. 
That is, from the general theory developed in Secs. 
\ref{switching scheme} and \ref{cv cluster state}, now we know 
that any approximate Gaussian cluster state can be 
deterministically generated in atomic ensembles trapped in a 
single-mode cavity. 
Moreover, as demonstrated here, the appropriate switching law 
can be systematically constructed, once the target cluster state 
is specified.


\section{Perturbation to pure steady state}
\label{sec:perturbation}

In this section, we reconsider the two atomic ensembles discussed 
in Sec.~\ref{sec:epr}, taking into account some specific 
perturbations added to the system. 
Actually, an atomic ensemble constructed in a cavity often loses 
coherence due to spontaneous emission. 
In addition to this kind of quantum effect, in practice any system 
contains some parameter uncertainties, which can bring a serious 
loss of coherence as well when aiming to dissipatively generate 
a pure state. 
In the case of an atomic ensemble, the number of atoms is usually 
never determined exactly. 
We take these two typical perturbations and evaluate how much 
the steady state is affected by these losses. 
In particular, we find the optimal squeezing level and the cavity 
damping rate that maximize the entanglement. 
Although these investigations do not straightforwardly provide 
new insight into the quasilocality discussed throughout this 
paper, the result will clarify a merit of enlarging the system 
rather than focusing only on the target system obtained by 
adiabatic elimination.


\subsection{Decoherence effect}

\begin{figure}[tb]
 \begin{center}
  \includegraphics[width=6cm]{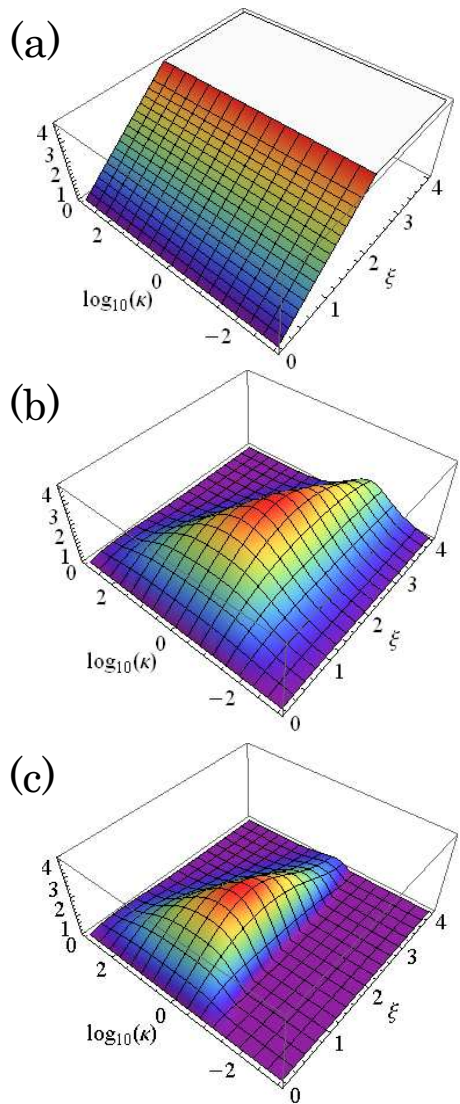}
 \end{center} 
 \caption{
 Logarithmic negativity $E_N$ versus the squeezing level 
 $\xi=\tanh^{-1}(r)$ and the cavity damping rate $\kappa$ 
 for the cases 
 (a) $\gamma=0$, 
 (b) $\gamma=0.01$, and 
 (c) $(\gamma, \epsilon)=(0.01, \sqrt{1.1})$. 
}
 \label{fig:neg}
\end{figure}

First we assume that the atomic ensembles are subjected to 
a loss of atomic coherence, in which case the corresponding 
coupling operator is represented by $L=\sqrt{\gamma}[a_1,a_2]^\top$,  
with $\gamma$ the decoherence rate. 
The coefficient matrix of the driving term of the system, 
given in Eq.~\eqref{A B matrices}, is then changed to 
\begin{align*}
    A' = \begin{bmatrix} 
            0 & \Sigma_2 \bar{C}^\top \\ 
            \Sigma_2 \bar{C} & -\kappa I_4/2 
         \end{bmatrix} 
        +\begin{bmatrix} 
            -\gamma I_4/2 & 0 \\ 
            0 & 0 \end{bmatrix}
        =A+\Delta A, 
\end{align*}
where $\bar{C}$ is defined via Eq.~\eqref{cepr}. 
The eigenvalues of $A$ are given by 
\begin{align*}
     \lambda_\pm 
        = -\frac{\kappa}{4} 
            \pm \frac{1}{4}\sqrt{\kappa^2 -16\mu^2 (1-r^2)}.
\end{align*}
A notable point is that, when large squeezing is introduced, 
i.e., $r\approx1$, the eigenvalue $\lambda_+$ approaches zero, 
even in the case $\gamma \ll \mu, \kappa$. 
This loss of stability of the system implies that a desirable 
convergence of the state is prevented. 
Thus the perturbation $\Delta A$ is not negligible, particularly 
when aiming to generate a large entangled state. 
Now the coefficient matrix of the diffusion term of the system 
is $B'=\text{diag}(\sqrt\gamma I_4,\sqrt\kappa I_4)$ and the 
Lyapunov equation $A'V+VA'\mbox{}^\top+B'B'\mbox{}^\top/2=0$ 
has the following explicit solution:
\begin{align*}
     V_1=\mathrm{diag}
           \left( \begin{bmatrix} 
                    rc+1/2 & -c \\ 
                    -c & rc+1/2 
                  \end{bmatrix}, 
                  \begin{bmatrix} 
                    rc+1/2 & c \\ 
                    c & rc+1/2 
                  \end{bmatrix} \right), 
\end{align*}
where $c=4r\kappa/(\kappa + \gamma)\{\kappa \gamma +4(1-r^2)\}$. 
Here, for simplicity, we have replaced $\kappa/\mu$ and $\gamma/\mu$ 
by $\kappa$ and $\gamma$, respectively. 
The entanglement of this two-mode Gaussian state can be quantified 
by the logarithmic negativity \cite{Vidal} and is given by
\begin{align*}
    & E_N=\max\{0,-\log(2\nu)\}, 
\\ \nonumber
    & \nu = \frac{\kappa \gamma^2 + 4\kappa (1-r)^2 
                      + \gamma [\kappa^2 +4(1-r^2)]}
                 {2(\kappa +\gamma)[\kappa \gamma +4(1-r^2)]}. 
\nonumber
\end{align*}
Figure \ref{fig:neg} shows $E_N$ versus the squeezing level 
$\xi=\tanh^{-1}(r)$ and the cavity damping rate $\kappa$, 
for the cases where the decoherence rate $\gamma$ takes 
the values (a) $\gamma=0$ and (b) $\gamma=0.01$. 
In the former case (a), since there is no decohering effect, the 
steady state is pure with maximal entanglement $E_N=2\xi$ without 
respect to $\kappa$; i.e., large squeezing directly means large 
entanglement. 
However, once the system is subjected to the decoherence, 
the steady-state entanglement drastically degrades, 
even for a very small decoherence rate $\gamma=0.01$, as depicted 
in Fig.~\ref{fig:neg}~(b). 
In particular, it is apparent from the figure that large 
squeezing brings about a large loss of entanglement, which is 
yet consistent with the fact that a large squeezed state is 
usually very fragile to decoherence. 
It is also reasonable that now $\kappa$ affects $E_N$ in two ways: 
Large $\kappa$ induces rapid leaking of the photons through the 
cavity light field, while small $\kappa$ means that the cavity 
mode interacts with the atomic ensembles many times, implying 
many emissions of the photons into the cavity field. 
Consequently, there exists an optimal set of the parameters 
$\kappa$ and $r=\tanh(\xi)$ that maximizes $E_N$: 
\begin{align}
\label{optimal parameters}
    & \kappa_\star =2\sqrt{1-r_\star^2}, 
\nonumber \\
    & r_\star =\frac{1}{2(\gamma^2+1)}
          \left( 2+d+\frac{\gamma^2 (\gamma^2 -3)}{d} \right), 
\end{align}
where 
$d=[-\gamma^6+5 \gamma^4-2 \gamma^2
+ (\gamma^2+1)\sqrt{-\gamma^4 (\gamma^2-4) }]^{1/3}$. 
Particularly in the case $\gamma=0.01$, these optimal values are 
given by $r_\star \approx 0.97$ and $\kappa_\star \approx 0.52$, 
and $E_N$ then takes the value of about $2.91$.

{\it Remark 4}. 
It should be pointed out that the above optimal parameters 
are out of the range where the auxiliary cavity mode can be 
adiabatically eliminated; 
that is, as discussed in Remark~1 in Sec.~\ref{sec:extended}, 
the auxiliary system can be adiabatically eliminated only when 
$\kappa$ is sufficiently large, but clearly in this case $E_N$ 
goes down to zero. 
In this sense, the system studied here provides an example 
where the extended system is really robust against decoherence 
compared to only the target system obtained through adiabatic 
elimination.


\subsection{Parameter uncertainty}

Since the condition imposing the system to have a pure steady state 
is described by a set of algebraic equations, it is easily violated 
by some parameter changes. 
In the case of the atomic system under consideration, the numbers 
of atoms of each ensemble must be exactly the same, i.e., $N_1=N_2$, 
but it is fairly unrealistic. 
Hence let us examine here how much the difference between $N_1$ and $N_2$ 
affects the entanglement of the steady state. 
We particularly set $\epsilon=\sqrt{N_2/N_1}=\sqrt{1.1}$; as usual, 
the number of trapped atoms is of order $10^6$, and such relatively 
large uncertainty (10$\%$) can actually happen. 
Figure \ref{fig:neg}-(c) shows the logarithmic negativity $E_N$, 
where the decoherence due to the spontaneous emission discussed in 
the preceding section is additionally taken into account. 
As expected, further degradation of the entanglement is 
observed and in almost all ranges of the parameters $\xi$ and 
$\kappa$ the state is no longer entangled. 
Nevertheless, surprisingly, it is not a uniform degradation; 
actually, as in the previous case, by engineering the system 
with optimal parameters 
$(\kappa_\star^\epsilon, r_\star^\epsilon)$, we obtain a steady 
state that still has a relatively large entanglement of $E_N=2.41$, 
which is only a 17$\%$ loss of entanglement compared to the ideal 
value of $E_N=2.91$, where there is no uncertainty (i.e., $\epsilon=1$). 
In other words, by constructing the system with these parameters 
$(\kappa_\star^\epsilon, r_\star^\epsilon)$, we can guarantee the 
entangled state with at least $E_N=2.41$, against the uncertainty 
of the difference of the number of atoms up to 10$\%$. 
This robustness property indicates the possible effectiveness of 
the dissipation-based method for state preparation even in a 
realistic situation, as long as the system parameters are 
appropriately determined. 
\\


\section{Conclusion}

The main results of this paper are twofold: 
First, we have shown that an arbitrary Gaussian pure state can 
be deterministically generated via the local dissipative environment, 
by constructing an appropriate auxiliary system. 
Second, we have shown that, even when only a single-mode auxiliary 
system is available, a well-tuned switching scheme allows us to 
stabilize any approximate Gaussian cluster state in a dissipative 
way. 
The former is a generalization of the scheme proposed in 
Ref. \cite{Parkins} that yields a dissipation-induced two-mode 
squeezed state, while the latter is that of Ref. \cite{Li2009} 
where deterministic generation of several four-mode cluster 
states was demonstrated.

The essential mechanism for bringing quasilocality to the system 
considered in this paper is that, for a Gaussian system, any 
interaction Hamiltonian is always a sum of two-body 
(hence quasilocal) Hamiltonians. 
This implies that as long as this kind of interaction Hamiltonian 
is taken when constructing an auxiliary system, any non-Gaussian 
(target) system couples to the environment quasilocally. 
This would be an interesting approach to explore a general method 
of constructing a desired quasilocal dissipative environment for 
general non-Gaussian systems.



\end{document}